\documentclass[%
reprint,
superscriptaddress,
 amsmath,amssymb,
 aps,
floatfix,
]{revtex4-2}

\usepackage{graphicx}
\usepackage{dcolumn}
\usepackage{bm}
\usepackage{appendix}

\usepackage{color}
\usepackage{physics}

\DeclareRobustCommand{\erase}{\bgroup\markoverwith{\textcolor{red}{\rule[.5ex]{2pt}{0.4pt}}}\ULon}

\begin{document}
\preprint{APS/123-QED}

\title{Transverse spin texture in optical non-Hermitian skin modes}

\author{Naoki Ichiji}
\affiliation{National Institute for Materials Science, 1-1 Namiki, Tsukuba, Ibaraki 305-0044, Japan}
\affiliation{Institute of Industrial Science, The University of Tokyo, 4-6-1 Komaba, Meguro-Ku, Tokyo 153-8505, Japan}
\author{Issei Takeda}
\affiliation{Department of Physics, Institute of Science Tokyo, 2-12-1 Ookayama, Meguro-ku, Tokyo 152-8550, Japan}
\affiliation{NTT Nanophotonics Center, NTT Inc. 3-1 Morinosato Wakamiya, Atsugi-shi, Kanagawa 243-0198, Japan}
\author{Taiki Yoda}
\affiliation{Department of Physics, Institute of Science Tokyo, 2-12-1 Ookayama, Meguro-ku, Tokyo 152-8550, Japan}
\author{Yuto Moritake}
\affiliation{Department of Physics, Institute of Science Tokyo, 2-12-1 Ookayama, Meguro-ku, Tokyo 152-8550, Japan}
\affiliation{Institute of Industrial Science, The University of Tokyo, 4-6-1 Komaba, Meguro-Ku, Tokyo 153-8505, Japan}
\author{Masaya Notomi}
\affiliation{Department of Physics, Institute of Science Tokyo, 2-12-1 Ookayama, Meguro-ku, Tokyo 152-8550, Japan}
\affiliation{Basic Research Laboratories, NTT Inc. 3-1 Morinosato Wakamiya, Atsugi-shi, Kanagawa 243-0198, Japan}
\affiliation{NTT Nanophotonics Center, NTT Inc. 3-1 Morinosato Wakamiya, Atsugi-shi, Kanagawa 243-0198, Japan}
\author{Satoshi Ashihara}
\affiliation{Institute of Industrial Science, The University of Tokyo, 4-6-1 Komaba, Meguro-Ku, Tokyo 153-8505, Japan}

\begin{abstract}
In structured electromagnetic fields, polarization textures are often closely linked to the spatial variation of the energy flow. However, this familiar picture has been established mainly for lossless and isotropic settings, and concrete examples showing how it is modified in media with gain and loss remain limited.
Here, we demonstrate that optical skin modes associated with the non-Hermitian skin effect (NHSE) carry a finite transverse circular-polarization texture and further show that the accompanying in-plane electric-field spin texture deviates from the familiar lossless spin-flow picture.
Using exact TE mode solutions, we separate the common exponential skin envelope from the oscillatory component. 
This decomposition shows that the circular-polarization texture is not generated by the skin envelope itself but by the oscillatory interference component modified by non-Hermiticity. It also reveals a handedness bias and a reshaped spatial relation between circularity and intensity.
Finite-element calculations confirm that these features remain robust in loss-biased anisotropic media. These results show that gain and loss provide additional freedom for engineering electric-field spin textures beyond conventional lossless photonic settings.
\end{abstract}

\maketitle
\section{Introduction}
Electromagnetic fields with spatial structure are now broadly discussed in the context of structured light, where spatial control of propagation and polarization has attracted wide interest~\cite{Roadmap17,Forbes21NatP}.
A key aspect of such fields is that the local polarization state can vary in space, allowing the associated optical spin to acquire three-dimensional orientations not fixed by a single propagation direction.
In particular, structured fields can support spin components orthogonal to the propagation direction, commonly referred to as transverse spin angular momentum (t-SAM).
This spatial freedom of polarization and spin has motivated extensive studies on the control of three-dimensional polarization textures and topological spin textures in structured electromagnetic fields~\cite{Rao24JOSAA,Lin21PRR,Tsesses18Sci,Du19NatPhys,Yanan20Nature}.

\begin{figure}[b!]
  \begin{center}
  \includegraphics[width=0.48\textwidth]{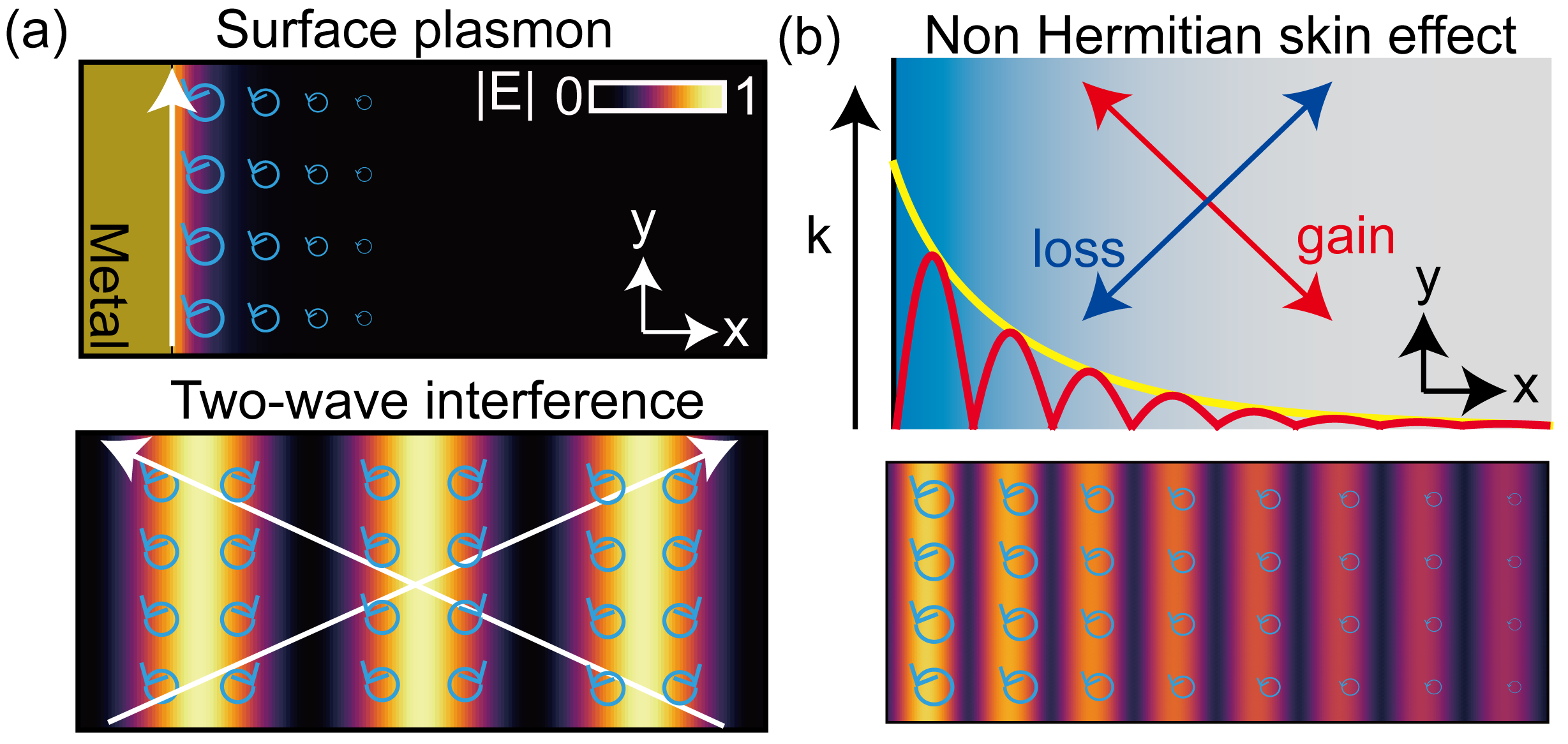}
  \end{center}
    \caption{(a) Schematic electric-field intensity maps and local circular-polarization states in representative structured fields: a surface-plasmon and a two-wave-interference field. Blue arrows indicate the local handedness, and white arrows indicate representative wave vectors.
    (b) Schematic of the optical non-Hermitian skin effect in an MT-symmetric anisotropic medium with balanced gain and loss along the principal axes. Yellow and red curves indicate the exponential envelope and oscillatory field profile of a representative skin mode, and the lower panel shows the corresponding skin-mode texture.}
  \label{Fig:concept}
\end{figure}

A canonical example is provided by evanescent waves, where the spin lies perpendicular to both the propagation and decay directions, and its sign is locked to the propagation direction~\cite{Bliokh12PRA,Nori14NCom} (Fig.~\ref{Fig:concept}(a)).
Related transverse-spin textures were later clarified in two-wave interference fields~\cite{Neugebauer15PRL,Nori15PRX,Bliokh15PRR}, in which the local handedness oscillates with the spatial modulation of the energy flow (Fig.~\ref{Fig:concept}(a)).
These examples revealed a close connection between transverse spin and the spatial variation of the energy flow.
In idealized lossless and isotropic formulations, this connection can be expressed more explicitly through spin-momentum equations.
For evanescent and guided waves, for example, transverse spin $\mathbf{S}_{\perp}$ is related to the curl of the energy flow $\mathbf{P}$, $\mathbf{S}_{\perp}\propto\nabla\times\mathbf{P}$~\cite{Shi21PNAS}.
This relation indicates that transverse spin reflects the spatial variation of the Poynting vector, and becomes pronounced where the energy flow varies strongly in space.

This spin-flow-based picture has since been recognized and reviewed as a broadly applicable framework for structured electromagnetic fields~\cite{Shi21NP}.
It is now used to describe polarization and spin textures in a wide range of systems beyond simple propagating waves, including waveguide evanescent modes~\cite{Shao18NC}, tightly focused beams~\cite{Winnerl12NJP,Aiello15NP}, and rotating near fields~\cite{Triolo17ACSP,Fu24NL,Ichiji25NL}.
These polarization and spin textures provide a basis for polarization- and spin-sensitive light--matter interactions, including optical forces and torques~\cite{Yuzhi22SciAd,Lu23PRL}, as well as spin transport in matter~\cite{Oue20PRB,Bekshaev22JOSAB}.

\newpage
However, most established spin-flow-based descriptions of polarization and spin textures have been formulated for lossless, typically isotropic, media.
Once gain, loss, or anisotropic dissipation becomes intrinsic to the mode-formation mechanism, it is no longer obvious that the familiar lossless picture remains valid in its standard form.
More specifically, concrete modal examples showing how the usual spin-flow correspondence is modified in such settings remain limited.

As a concrete example of such a setting, we analyze optical skin modes associated with the non-Hermitian skin effect (NHSE).
Non-Hermitian systems, described by effective Hamiltonians with gain and loss, can exhibit the NHSE, a localization phenomenon in which eigenmodes under open boundary conditions accumulate at system boundaries~\cite{Yao18PRL,Okuma20PRL}.
In reciprocal two-dimensional optical continuous media, recent studies have shown that such skin modes appear as surface-wave-like propagating eigenmodes, where standing-wave-like modes acquire an exponential envelope and localize toward one boundary while propagating along it~\cite{Yoda25PRR,Takeda25PRA} (Fig.~\ref{Fig:concept}(b)).
Because these modes combine surface-wave-like localization with oscillatory interference structure, they provide a natural setting for examining how the familiar spin textures of evanescent waves and two-wave interference fields are modified in a non-Hermitian medium.

We show that optical NHSE skin modes exhibit a finite transverse circular-polarization texture of the in-plane electric field, demonstrating the presence of an electric-field spin component perpendicular to the propagation direction.
Despite their surface-wave-like localization, this spin texture originates not from the exponential skin envelope but from the oscillatory interference component of the mode.
At the same time, non-Hermiticity modifies this interference-derived texture, making it distinct from both conventional evanescent-wave and lossless two-wave-interference spin textures.
Finally, by extending the analysis to loss-biased anisotropic media, we confirm that these characteristic features remain qualitatively robust beyond the ideal mirror-time (MT)-symmetric limit.

In characterizing these spin textures, we do not directly evaluate the full optical spin-angular-momentum density in the material medium, whose rigorous formulation would require a consistent treatment of the material response, including dispersion, anisotropy, and gain and loss~\cite{Bliokh17PRL,Bliokh17NJP,Philbin12PRA,Yamashita11OptCom,Hashiyada25PRR}.
Instead, we use Stokes parameters constructed from the complex electric-field eigenmodes as pointwise descriptors of the local polarization state.
In particular, the third Stokes parameter $S_3$ quantifies the in-plane electric-field circularity and provides a physically meaningful descriptor of the electric-field contribution associated with transverse spin.
We therefore refer to the spatial distribution of this $S_3$-defined circularity as the transverse spin texture of the NHSE skin mode.

\section{TE skin modes in an MT-symmetric medium}
\subsection{Geometry and analytical mode formulation}
Before analyzing the polarization texture, we briefly summarize the analytical mode solution for optical NHSE skin modes in a finite-width MT-symmetric medium with balanced gain and loss.
The non-Hermitian skin effect was originally discussed mainly in discrete systems, where eigenmodes under open boundary conditions accumulate at system boundaries~\cite{Yao18PRL, Yokomizo19PRL, Okuma20PRL}.
More recently, optical NHSE has been shown to arise in reciprocal continuous anisotropic media, where confined modes acquire a common exponential envelope and become localized toward one boundary while retaining a standing-wave-like oscillation~\cite{Yoda25PRR,Takeda25PRA}.

\begin{figure*}[t]
\centering
\includegraphics[width=0.95\textwidth]{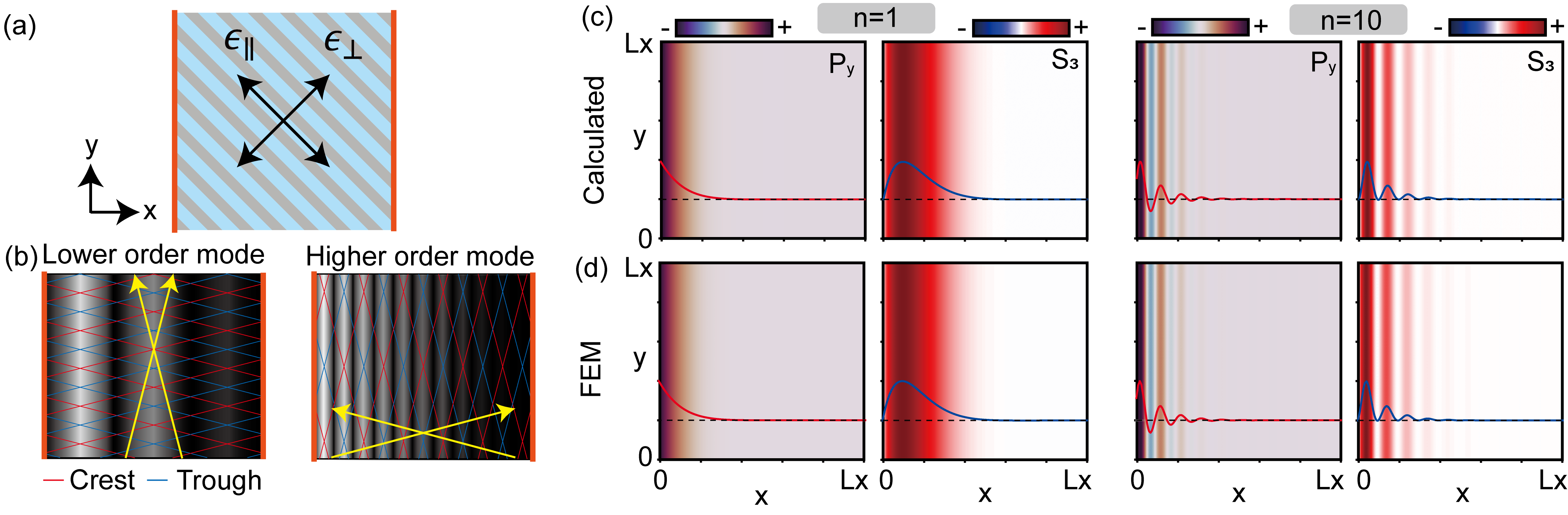}
\caption{Energy-flow and polarization textures in an MT-symmetric medium. (a) Schematic of the MT-symmetric anisotropic medium. (b) Schematic field structures of low- and high-order skin modes. The yellow arrows indicate the wave vectors of the constituent plane-wave components, and the red and blue lines indicate the crests and troughs of these components.} (c,d) Axial energy-flow density $P_y(x,y)$ and third Stokes parameter $S_3(x,y)$ for the fundamental ($n=1$) and higher-order ($n=10$) modes. The calculations were performed for $\varepsilon'=1$, $\varepsilon''=1$, $k_y=0.5\times 2\pi/\lambda_{\rm ref}$, $\lambda_{\rm ref}=5~\mu\mathrm{m}$, and $L_x=10~\mu\mathrm{m}$. (c) Distributions reconstructed from the closed-form analytical solution. (d) Corresponding FEM results. The line plots indicate representative cross sections extracted from the corresponding two-dimensional maps.
\label{Fig:PySz_compare}
\end{figure*}

Here, we consider a TE mode in a finite-width waveguide bounded by perfect-electric-conductor (PEC) walls at $x=0$ and $x=L_x$, with propagation constant $k_y$ along the $y$ direction, as in Refs.~\cite{Yoda25PRR,Takeda25PRA}.
As shown in Fig.~\ref{Fig:PySz_compare}(a), the medium has principal permittivities $\varepsilon_{\parallel}$ and $\varepsilon_{\perp}$, and its principal axes are tilted by $45^\circ$ with respect to the coordinates $(x,y)$.
The in-plane permittivity tensor is then written as
\begin{equation}
\varepsilon=
\begin{pmatrix}
\varepsilon_{xx} & \varepsilon_{xy}\\
\varepsilon_{yx} & \varepsilon_{yy}
\end{pmatrix}
=
\begin{pmatrix}
(\varepsilon_{\parallel}+\varepsilon_{\perp})/2 & (\varepsilon_{\parallel}-\varepsilon_{\perp})/2\\
(\varepsilon_{\parallel}-\varepsilon_{\perp})/2 & (\varepsilon_{\parallel}+\varepsilon_{\perp})/2
\end{pmatrix}.
\label{eq:eps_rot45_general}
\end{equation}

The in-plane permittivity tensor is denoted by $\varepsilon_{ij}$ with inverse
$\eta_{ij}=(\varepsilon^{-1})_{ij}$, and we define
\begin{equation}
q=\frac{\varepsilon_{xy}+\varepsilon_{yx}}{2\varepsilon_{xx}},\qquad
\beta = \frac{\pi n}{L_x},\qquad
k_\pm = \pm \beta - q k_y,
\end{equation}
together with
\begin{align}
Z_x^\pm &= \frac{1}{\omega\varepsilon_0}\bigl(-k_y\eta_{yx} + k_\pm\eta_{yy}\bigr),\label{eq:Zx_def}\\
Z_y^\pm &= \frac{1}{\omega\varepsilon_0}\bigl(k_y\eta_{xx} - k_\pm\eta_{xy}\bigr).\label{eq:Zy_def}
\end{align}
where $\omega$ is the angular frequency and $\varepsilon_0$ is the vacuum permittivity, and $k_\pm$ denote the transverse wave numbers of the two constituent plane-wave components forming the confined transverse mode.

For a given integer $n$, the electric and magnetic fields of the $n$-th TE mode can be written compactly as
\begin{align}
E_y(x) &= e^{i q k_y x}\, \sin\!\left(\beta x\right),\label{eq:ori_Ey}\\[2mm]
H_z(x) &= \frac{1}{2} e^{i q k_y x}\bigl[ B_C \cos(\beta x) + B_S \sin(\beta x)\bigr],\label{eq:ori_Hz}\\[2mm]
E_x(x) &= -\frac{1}{2} e^{i q k_y x}\bigl[ A_C \cos(\beta x) + A_S \sin(\beta x)\bigr]\label{eq:ori_Ex},
\end{align}

where the coefficients are defined by
\begin{align}
A_C &= i\left(\frac{Z_y^+}{Z_x^+} - \frac{Z_y^-}{Z_x^-}\right), &
A_S &= \left(\frac{Z_y^+}{Z_x^+} + \frac{Z_y^-}{Z_x^-}\right), \\
B_C &= i\left(\frac{1}{Z_x^+} - \frac{1}{Z_x^-}\right), &
B_S &= \left(\frac{1}{Z_x^+} + \frac{1}{Z_x^-}\right).
\end{align}

The field profiles in Eqs.~(\ref{eq:ori_Ey})--(\ref{eq:ori_Ex}) are thus described by standing-wave-like oscillatory terms, represented by $\sin(\beta x)$ and $\cos(\beta x)$ and determined by the finite-width confinement and boundary conditions. Their relative weights and the common complex envelope factor $e^{iqk_yx}$ are set by the anisotropic permittivity tensor.
The transverse modal profile can therefore be viewed as an oscillatory interference structure modulated by a common complex envelope.
Here, the mode index $n$ effectively determines the transverse wavenumber $\beta$.
As schematically shown in Fig.~\ref{Fig:PySz_compare}(b), higher-order modes correspond to a larger transverse wave number $\beta$ and thus to constituent waves that are more oblique with respect to the $y$ direction.

These expressions are valid for arbitrary in-plane anisotropic permittivity tensors $\varepsilon_{ij}$.
We now specialize to an MT-symmetric non-Hermitian medium~\cite{Yoda25PRR,Takeda25PRA},
\begin{equation}
\varepsilon
=
\begin{pmatrix}
\varepsilon' & i \varepsilon'' \\
i \varepsilon'' & \varepsilon'
\end{pmatrix},
\label{eq:eps_MT}
\end{equation}
and introduce the dimensionless parameter $\alpha=\varepsilon''/\varepsilon'$.
In this case, $q=i\alpha$, so that the common factor becomes
\begin{equation}
e^{iqk_yx}=e^{-\alpha k_y x}.
\end{equation}
This expression provides a purely exponential envelope along the $x$-direction and gives rise to skin localization under open boundary conditions. 
In this picture, the MT-symmetric skin mode is understood as a confined oscillatory mode whose amplitude is exponentially biased toward one boundary by the non-Hermitian anisotropy.

\subsection{Poynting vector and Stokes parameter}

Based on the above formulation, we first analyze the TE skin modes of an MT-symmetric finite-width medium using the closed-form solutions in Eqs.~(\ref{eq:ori_Ey})--(\ref{eq:ori_Ex}).
From the resulting complex eigenfields, we evaluate the time-averaged energy-flow density
\begin{equation}
\mathbf P(\mathbf r)=\frac12\,\mathrm{Re}\!\left[\mathbf E(\mathbf r)\times \mathbf H^*(\mathbf r)\right],
\label{eq:P_general}
\end{equation}
and characterize the local circular polarization in the $x$-$y$ plane through the third Stokes parameter
\begin{equation}
S_3(\mathbf r)=2\,\mathrm{Im}\!\left[E_x^*(\mathbf r)E_y(\mathbf r)\right].
\label{eq:stokes_s3}
\end{equation}
Here, $S_3$ denotes the handedness of the local in-plane circular polarization defined from the field components $E_x$ and $E_y$.

Figure~\ref{Fig:PySz_compare} shows the spatial maps of $P_y(x,y)$ and $S_3(x,y)$ reconstructed from the analytic mode fields in Eqs.~(\ref{eq:ori_Ey})--(\ref{eq:ori_Ex}) for the fundamental ($n=1$) and higher-order ($n=10$) modes at the same $(\varepsilon',\varepsilon'',k_y,L_x)$.
We first focus on the fundamental mode ($n=1$).
The analytic result for $P_y(x,y)$ shows pronounced localization at the left edge, confirming the expected skin accumulation.
By contrast, $P_x$ vanishes identically in the analytic MT-symmetric solution (not shown).
This energy-flow structure closely resembles that of a conventional surface wave such as a plane SPP (see Appendix~\ref{app:spp_compare}), confirming that the NHSE skin mode is not merely a boundary-accumulated eigenfield, but an energy-carrying propagating mode.

At the same time, the corresponding polarization map exhibits a finite $S_3(x,y)$ localized in the same skin region.
This reveals that the optical skin mode is accompanied by a localized transverse circularly polarized state, namely a skin-localized in-plane rotation of the electric field, in analogy with other structured electromagnetic fields such as evanescent waves and two-wave interference fields.

However, a notable feature here is that this localized $S_3$ profile does not follow the t-SAM profile of a conventional evanescent wave, despite the surface-wave-like appearance of the mode.
As discussed in the Introduction, in representative lossless isotropic media the t-SAM is proportional to the spatial variation of the Poynting vector~\cite{Shi21PNAS,Shi21NP}.
For a plane SPP, this leads to a t-SAM profile that follows the evanescent decay of the energy flow, and because the transverse spin is purely electric~\cite{Ichiji25NL,Peng19NatP,Bliokh12PRA}, the third Stokes parameter $S_3$ likewise exhibits an exponentially decaying profile.
In the present NHSE skin mode, by contrast, although $P_y$ decays monotonically away from the interface, $S_3$ vanishes at $x=0$ and reaches its maximum at a finite distance from the boundary.
This behavior is clearly different from that of a conventional surface wave.

Higher-order modes reveal additional internal structure.
For $n=10$, the energy-flow density $P_y$ remains confined within the same exponential envelope but develops pronounced oscillatory modulations across $x$.
We also find that $P_y$ exhibits local negative regions within the skin-localized profile, so that its sign changes periodically along the $x$ direction rather than remaining unidirectional.
The third Stokes parameter $S_3$ also exhibits an oscillatory modulation within the same localized region, although its negative part is strongly suppressed and is barely visible. 
Thus, the $S_3$ profile retains the oscillatory structure while remaining nearly unidirectional in sign.

This behavior is also distinct from that of a conventional two-wave-interference field.
If the usual spin-flow relation were directly applicable to the full skin-localized profile, the spatial modulation of $P_y$ would lead to a sign-changing $S_3$ profile, as in lossless two-wave interference fields.
Such a sign-changing circular-polarization profile is not observed here; instead, the oscillatory $S_3$ profile is biased toward one handedness.
These features indicate that the polarization texture of the NHSE skin mode cannot be captured by the conventional spin-flow relation in their standard form.

For comparison, we also performed finite-element eigenmode calculations using COMSOL Multiphysics for the same geometry and material parameters.
Specifically, we considered a TE eigenmode problem with field dependence $\exp(ik_y y-i\omega t)$ at a prescribed propagation constant $k_y$.
PEC boundary conditions were imposed on both sidewalls, and the $y$ direction was treated with a Floquet periodic condition fixing $k_y$. Unless otherwise noted, the numerical setup follows Refs.~\cite{Yoda25PRR,Takeda25PRA}. The FEM results show good agreement with the analytic profiles for both $P_y$ and $S_3$ across all mode indices examined (Fig.~\ref{Fig:PySz_compare}(d)), confirming that the closed-form analytical solution captures the essential energy-flow and polarization textures of the NHSE skin modes.

\section{Field decomposition and polarization textures}
\subsection{Coefficient decomposition and mode-index dependence}
To clarify the origin of these polarization textures, we analyze the MT-symmetric skin modes in more detail by decomposing the field components and deriving closed-form expressions for the relevant observables.
In the MT-symmetric case, the general solution in Eqs.~(\ref{eq:ori_Ey})--(\ref{eq:ori_Ex}) can be written in a simpler form by substituting the permittivity tensor in Eq.~(\ref{eq:eps_MT}), yielding the corresponding field coefficients and mode profiles.

Substituting Eq.~(\ref{eq:eps_MT}) into Eqs.~(\ref{eq:Zx_def}) and~(\ref{eq:Zy_def}), we find that $Z_x^\pm$ and $Z_y^\pm$ can be written as
\begin{equation}
Z_{x}^{\pm} = \pm u(n),\qquad
Z_{y}^{\pm} = v \pm i w(n).
\label{eq:Z_MT_compact}
\end{equation}

Here $u(n)$, $v$, and $w(n)$ are real coefficients defined by
\begin{align}
u(n) &= \frac{\beta \varepsilon'}{\omega\varepsilon_0\bigl(\varepsilon'^2+\varepsilon''^2\bigr)},
\label{eq:coef_u}
\\[1mm]
v    &= \frac{k_y}{\omega\varepsilon_0\varepsilon'},
\label{eq:coef_v}
\\[1mm]
w(n) &= \frac{\beta \varepsilon''}{\omega\varepsilon_0\bigl(\varepsilon'^2+\varepsilon''^2\bigr)}.
\label{eq:coef_w}
\end{align}

Here $u(n)$ and $w(n)$ grow linearly with the mode index via $\beta\propto n$, while $v$ is independent of $n$. Using Eq.~\eqref{eq:Z_MT_compact}, the field coefficients in Eqs.~(\ref{eq:ori_Ey})--(\ref{eq:ori_Ex}) reduce to
\begin{align}
B_C &= \frac{2i}{u(n)},\qquad
B_S = 0,\nonumber\\
A_C &= \frac{2vi}{u(n)},\qquad
A_S = 2i\alpha.
\label{eq:AB_MT}
\end{align}

Equation~\eqref{eq:AB_MT} shows that $B_S$ vanishes identically and that $A_S$ is fixed solely by the non-Hermitian parameter $\alpha$, whereas both $A_C$ and $B_C$ are $n$-dependent and decrease with the mode index through $u(n)$.

Combining Eq.~\eqref{eq:AB_MT} with Eq.~\eqref{eq:ori_Ey}, and using $q=i\alpha$, we obtain the MT-symmetric skin-mode profiles
\begin{align}
E_y(x) &= e^{-\alpha k_y x}\,\sin(\beta x),
\label{eq:Ey_MT_final}
\\[1mm]
H_z(x) &= \frac{i}{2} e^{-\alpha k_y x}\,\frac{2}{u(n)}\cos(\beta x),
\label{eq:Hz_MT_final}
\\[1mm]
E_x(x) &= -\frac{i}{2} e^{-\alpha k_y x}
\left[
\frac{2 v}{u(n)}\cos(\beta x) + 2\alpha\sin(\beta x)
\right].
\label{eq:Ex_MT_final}
\end{align}

\begin{figure}[t]
\centering
\includegraphics[width=0.45\textwidth]{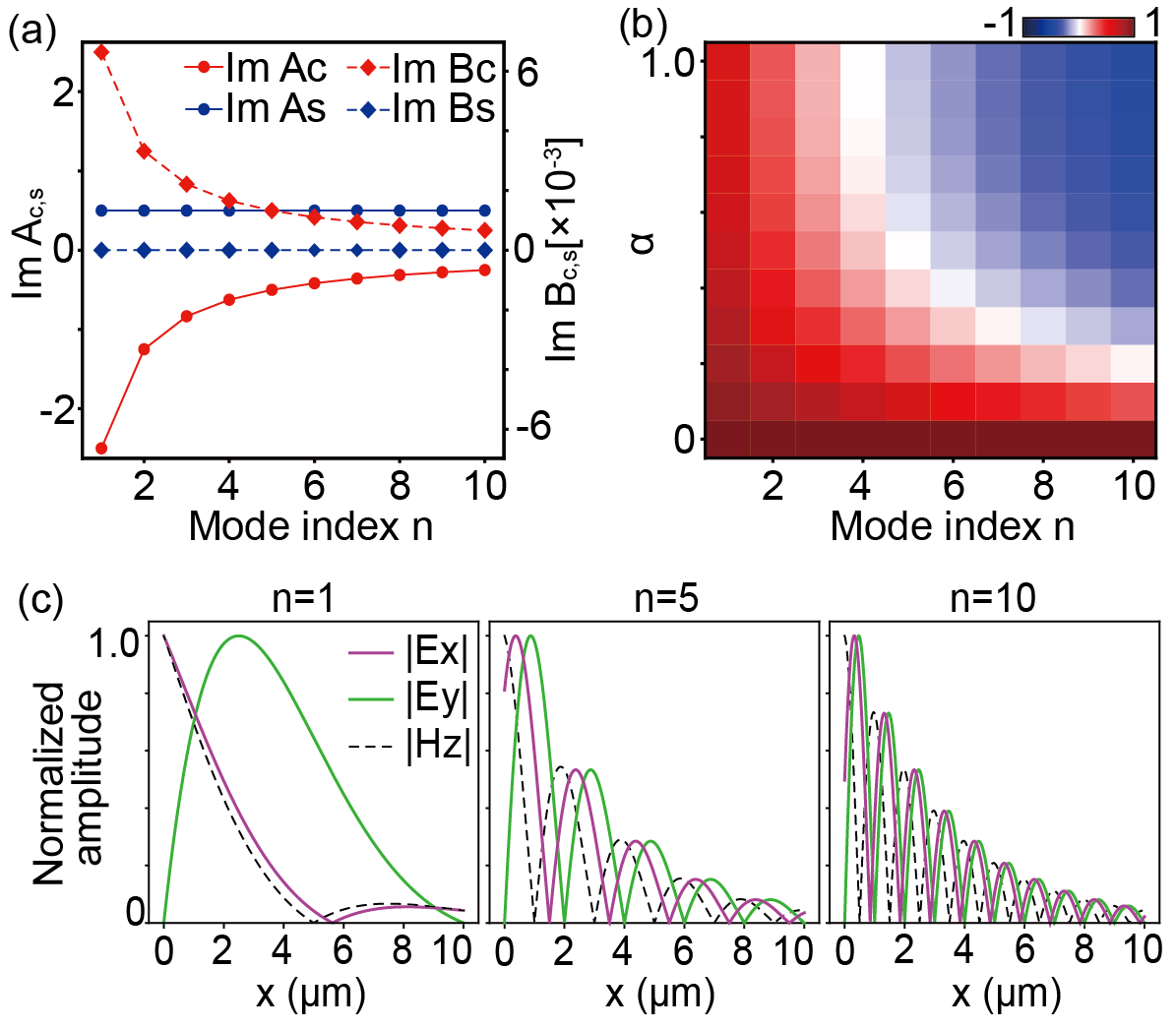}
\caption{Field structure of the MT-symmetric NHSE skin modes.
(a) Mode-index dependence of the coefficients $\Im A_C$, $\Im A_S$, $\Im B_C$, and $\Im B_S$ for a representative parameter set with $f=29.979~\mathrm{THz}$, $\alpha=0.5$, and $k_y=k_2\,2\pi/\lambda_{\rm ref}$ with $k_2=0.5$ and $\lambda_{\rm ref}=5~\mu\mathrm{m}$.
(b) Signed balance map of the cosine- and sine-like contributions in $E_x$ on the $(n,\alpha)$ plane at fixed $k_y=0.5\times 2\pi/\lambda_{\rm ref}$. (c) Normalized cross-sectional profiles of $|E_x(x)|$, $|E_y(x)|$, and $|H_z(x)|$ along $y=0$ for $n=1$, $5$, and $10$ in a finite-width medium with} $L_x=10~\mu\mathrm{m}$, calculated for the same parameter set as in panel (a). \label{Fig:EHfield}
\end{figure}

These expressions show that $E_y$ and $H_z$ retain pure sine- and cosine-like profiles, whereas only $E_x$ contains both cosine- and sine-like components. In this sense, the non-Hermitian anisotropy modifies the field structure across the waveguide width specifically through the mode-dependent reshaping of $E_x$.

These features can be interpreted in terms of the following physical picture.
The tangential component $E_y$ is directly constrained by the electric-wall boundary condition at the sidewalls, $E_y(x=0)=E_y(x=L_x)=0$, and therefore retains a pure sine profile. By contrast, the boundary condition does not directly prescribe the profile of $E_x$ in the same way as that of $E_y$, so $E_x$ is not restricted to a pure cosine profile. In the non-Hermitian anisotropic medium, the off-diagonal terms of the permittivity tensor couple the $E_x$ and $E_y$~\cite{Born99Book}, giving rise to an additional $E_y$-derived sine component in $E_x$.

Equation~(\ref{eq:Ex_MT_final}) shows that the sine term in $E_x$ is proportional to $\alpha$, indicating that the contribution arising from the off-diagonal anisotropy becomes stronger as $\alpha$ increases. The same equation also shows that the cosine term is weighted by $v/u(n)$, which decreases as the mode index $n$ increases. As schematically illustrated in Fig.~\ref{Fig:concept}(b), higher-order modes correspond to constituent waves that are more oblique to the propagation direction, so that the TE field becomes increasingly $E_y$-dominant. As a result, the sine component in $E_x$ induced by the off-diagonal anisotropy becomes relatively more prominent.
Finally, although $H_z$ is not constrained by the boundary condition in the same way as $E_y$, the MT-symmetric coefficient condition yields $B_S=0$, so that its sine component vanishes and a purely cosine-like profile is retained.

Figure~\ref{Fig:EHfield}(a) plots $\operatorname{Im}A_C$, $\operatorname{Im}A_S$, $\operatorname{Im}B_C$, and $\operatorname{Im}B_S$ as functions of the mode index $n$ for a representative parameter set, $\alpha=0.5$ and $k_y=0.5\times 2\pi/\lambda_{\mathrm{ref}}$. As expected from Eq.~\eqref{eq:AB_MT}, $B_S$ remains zero for all $n$, while $|B_C|$ decreases with increasing $n$. Importantly, $|A_C|$ decreases whereas $|A_S|$ stays constant, so that the relative balance between the cosine- and sine-like contributions in $E_x$ changes systematically with the mode index. Consequently, $E_x$ is cosine-like for lower-order modes, but becomes increasingly sine-like as the mode index increases.

To summarize the balance between the two contributions to $E_x$, Fig.~\ref{Fig:EHfield}(b) plots the signed balance parameter
\begin{equation}
D_A=\frac{|\operatorname{Im}A_C|-|\operatorname{Im}A_S|}{|\operatorname{Im}A_C|+|\operatorname{Im}A_S|},
\end{equation}
on the $(n,\alpha)$ plane at fixed $k_y$.
Here, positive and negative values of $D_A$ indicate cosine-like and sine-like dominance in $E_x$, respectively.
The map shows that the balance shifts toward the sine-like side as either the mode index $n$ or the non-Hermitian parameter $\alpha$ increases.
This follows directly from Eq.~(\ref{eq:Ex_MT_final}) because increasing $\alpha$ enhances the sine-like term proportional to $\alpha$, whereas increasing $n$ suppresses the cosine-like term through $v/u(n)\propto k_y/n$.

This trend is confirmed by the normalized amplitude profiles in Fig.~\ref{Fig:EHfield}(c), which show $|E_y|$, $|E_x|$, and $|H_z|$ along the $x$ direction at $y=0$ for representative modes, $n=1$, $n=5$, and $n=10$, calculated from Eqs.~(\ref{eq:Ey_MT_final})--(\ref{eq:Ex_MT_final}).

In all cases, the profiles exhibit the expected exponential envelope $\exp(-\alpha k_y x)$ together with the oscillation period set by $L_x/n$, confirming the standing-wave-like structure along $x$ within the skin-localized mode.
For the lower-order mode, $|E_x|$ more closely follows $|H_z|$ than $|E_y|$, whereas for the higher-order mode, the maxima and minima of $|E_x|$ become nearly aligned with those of $|E_y|$. This real-space change is consistent with the mode-dependent balance between the cosine- and sine-like contributions to $E_x$ discussed above.

\subsection{Closed-form expressions for $P_y$ and $S_3$}
The analysis above shows that the MT-symmetric NHSE skin modes possess a field structure distinct from both ordinary oscillatory interference fields and conventional surface waves. We now use this characteristic field structure to derive closed-form expressions for the axial energy flow $P_y(x)$ and the third Stokes parameter $S_3(x)$. 

Using Eqs.~(\ref{eq:Ey_MT_final})–(\ref{eq:Ex_MT_final}) together with Eqs.~(\ref{eq:P_general}) and~(\ref{eq:stokes_s3}), $P_{y}$ and $S_{3}$ are expressed as
\begin{align}
  P_y(x)
  = \frac{e^{-2\alpha k_y x}}{4u(n)^2}
     \Bigl[v\bigl(1+\cos 2\beta x\bigr) + \alpha u(n)\,\sin 2\beta x
     \Bigr],
\label{eq:Py_MT}
\\[1mm]
  S_3(x)
  = \frac{e^{-2\alpha k_y x}}{u(n)}
    \Bigl[
      \alpha u(n)\bigl(1-\cos 2\beta x\bigr) + v\sin 2\beta x
    \Bigr].
\label{eq:S3_MT}
\end{align}

In the Hermitian limit $\alpha\to 0$, Eqs.~(\ref{eq:Py_MT}) and~(\ref{eq:S3_MT}) reduce to
\begin{align}
  P_y(x) &= \frac{v}{4u(n)^2}\,\bigl[1+\cos 2\beta x\bigr],\\
  S_3(x) &= \frac{v}{u(n)}\,\sin 2\beta x.
\end{align}
In this limit, $\partial_x P_y$ is proportional to $S_3$, which reproduces the familiar two-wave-interference correspondence between the energy-flow gradient and the local circular-polarization texture~\cite{Shi21PNAS, Ichiji23PRA, Kihara25OptComun}.

By contrast, once non-Hermiticity is introduced, the full profiles of $P_y(x)$ and $S_3(x)$ no longer satisfy such a simple proportionality, as also confirmed by a direct comparison between $S_3$ and $\partial_x P_y$ (see Appendix~\ref{app:dPy_compare}). Nevertheless, the structure of Eqs.~(\ref{eq:Py_MT}) and~(\ref{eq:S3_MT}) suggests that a related correspondence may still survive at the level of the oscillatory part.
To disentangle the gain/loss-induced envelope from the underlying standing-wave-like interference, we extract the oscillatory parts from Eqs.~(\ref{eq:Py_MT}) and~(\ref{eq:S3_MT}) as
\begin{align}
  P_y^{\mathrm{osc}}(x)
  &= v\bigl(1+\cos 2\beta x\bigr) + \alpha u(n)\,\sin 2\beta x,
  \label{eq:Py_osc_def}\\[1mm]
  S_3^{\mathrm{osc}}(x)
  &= \alpha u(n)\bigl(1-\cos 2\beta x\bigr) + v\sin 2\beta x,
  \label{eq:S3_osc_def}
\end{align}
which yields
\begin{equation}
  S_3^\mathrm{osc}
  = -\,\frac{1}{2\beta}\,\frac{\partial P_y^\mathrm{osc}}{\partial x}
  + \alpha u(n).
  \label{eq:G_F_relation}
\end{equation}
Equation~(\ref{eq:G_F_relation}) shows that the oscillatory parts of $P_y$ and $S_3$ satisfy a simple differential relation, up to the constant offset term $\alpha u(n)$ induced by the non-Hermitian shift of $E_x$ in Eq.~(\ref{eq:Ex_MT_final}).

\begin{figure}[t]
\centering
\includegraphics[width=0.48\textwidth]{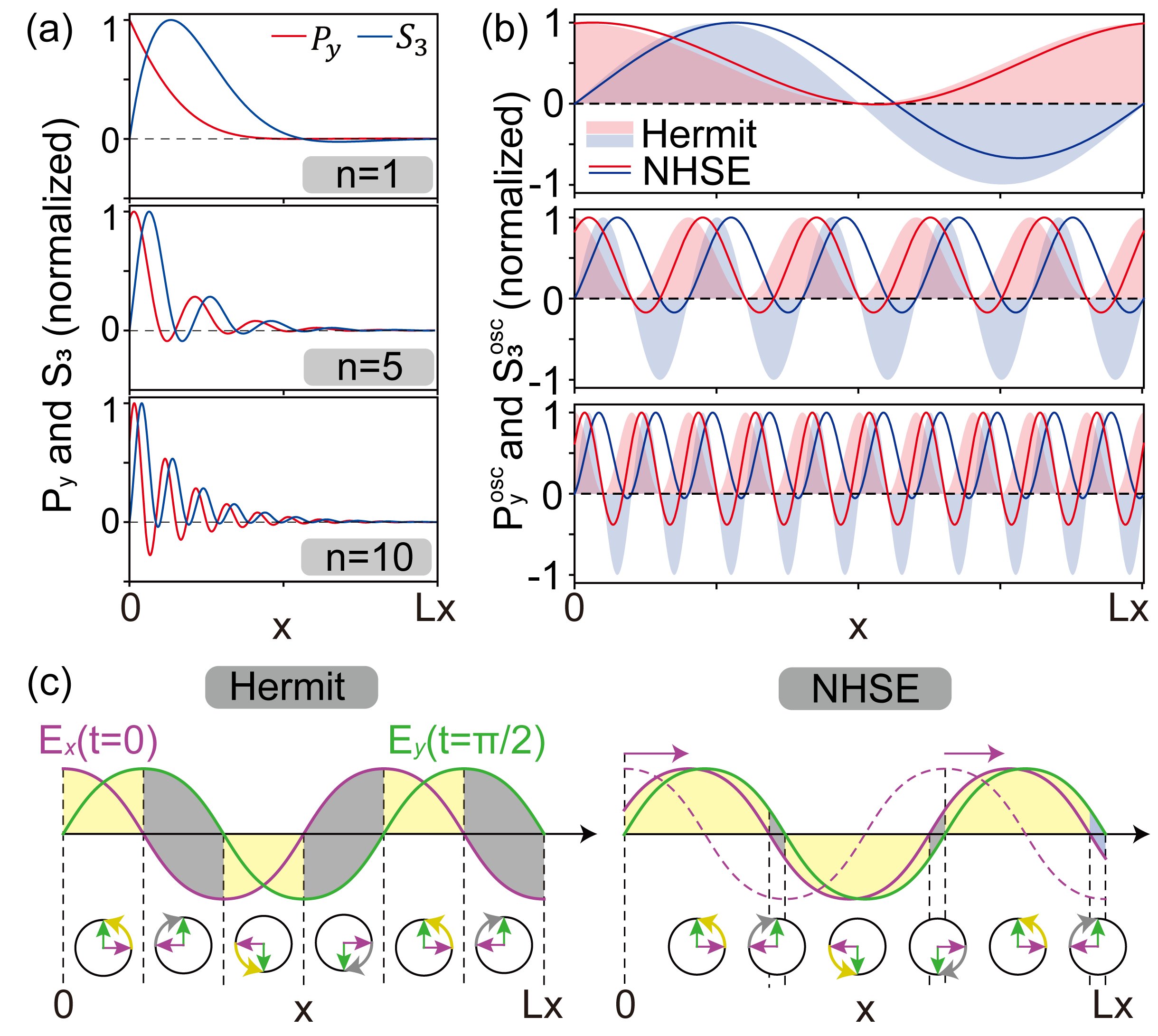}
\caption{Normalized axial energy flow $P_y (x)$ (blue) and Stokes parameter $S_3 (x)$ (red) of TE skin modes in an MT-symmetric medium, shown for mode indices $n=1$ (top), $n=5$ (middle), and $n=10$ (bottom), (a) Full profiles including the exponential skin envelope (near-edge region, $0\leq x\leq3~\mathrm{\mu m}$). (b) Normalized oscillatory components $P_y^{\mathrm{osc}} (x)$ and $S_3^{\mathrm{osc}} (x)$. The dashed lines show the Hermitian limit ($\alpha = 0$) for each case.
(c) Schematic comparison of the Hermitian and NHSE field structures in terms of the quadrature components $E_x(t=0)$ and $E_y(t=\pi/2)$ and the corresponding local polarization ellipses.}
\label{fig:PyS3_profiles}
\end{figure}

Figure~\ref{fig:PyS3_profiles} summarizes the energy-flow and polarization textures of the MT-symmetric skin modes.
Panel (a) shows the full near-edge profiles of $P_y(x)$ and $S_3(x)$ including the exponential envelope $e^{-2\alpha k_y x}$, whereas panel (b) shows the corresponding oscillatory parts $P_y^{\mathrm{osc}}(x)$ and $S_3^{\mathrm{osc}}(x)$ defined in Eqs.~(\ref{eq:Py_osc_def}) and~(\ref{eq:S3_osc_def}), with the Hermitian-limit curves ($\alpha=0$) overlaid as shaded regions.
The curves in panel (b) confirm the differential relation implied by Eq.~(\ref{eq:G_F_relation}) and show that the constant offset term $\alpha u(n)$ systematically biases $S_3^{\mathrm{osc}}(x)$ toward one handedness within the skin-localized mode.
In the Hermitian limit, corresponding to an ordinary two-wave interference field, the positive and negative parts of $S_3$ are distributed evenly over one transverse period, so that the net handedness averages out. In the NHSE case, by contrast, the offset breaks this balance and biases the oscillatory polarization texture toward one handedness.

Taken together, Eqs.~(\ref{eq:Py_MT})--(\ref{eq:G_F_relation}) and Fig.~\ref{fig:PyS3_profiles} indicate that the polarization texture is governed by the oscillatory interference terms, whereas the gain/loss-induced exponential factor does not itself generate an additional contribution to $S_3$ and merely acts as a common envelope multiplying both $P_y$ and $S_3$.

This distinction reflects the different origins of the two spatial structures.
The common exponential envelope originates neither from interference nor from boundary-induced evanescent decay as in SPPs, but from the direction-dependent gain and loss encoded in the anisotropic permittivity tensor.
A gain/loss-induced amplitude gradient can attenuate or amplify the field intensity without changing the local polarization state. 
The exponential envelope is therefore distinct from the spin-generating spatial structure in conventional lossless structured fields.
By contrast, the oscillatory parts arise from the same interference mechanism that governs ordinary two-wave-interference fields, now embedded in a gain/loss medium.
For this reason, they retain the differential relation between $P_y$ and $S_3$ familiar from the lossless case, while being modified by the offset term in Eq.~(\ref{eq:G_F_relation}).

The handedness bias can be understood from the lateral shift of $E_x$, as schematically illustrated in Fig.~\ref{fig:PyS3_profiles}(c).
Because $E_x$ and $E_y$ are in temporal quadrature, as seen from Eqs.~(\ref{eq:Ey_MT_final}) and~(\ref{eq:Ex_MT_final}), the in-plane circularity is finite everywhere except at points where either component is 0.
For a fixed quadrature order, the handedness alternates depending on whether $E_x$ and $E_y$ have the same sign or opposite signs in the transverse profile, equivalently according to the sign of $\mathrm{Im}(E_x^*E_y)$.
In the Hermitian limit, these regions are distributed evenly over one transverse period, so that the net circularity averages out.
When non-Hermiticity introduces a lateral shift in $E_x$, this balance is broken, leading to an unequal weighting of the two handednesses and hence to a constant offset in the oscillatory circular-polarization texture within the skin-localized region.

\begin{figure*}[t!]
\centering
\includegraphics[width=0.9\textwidth]{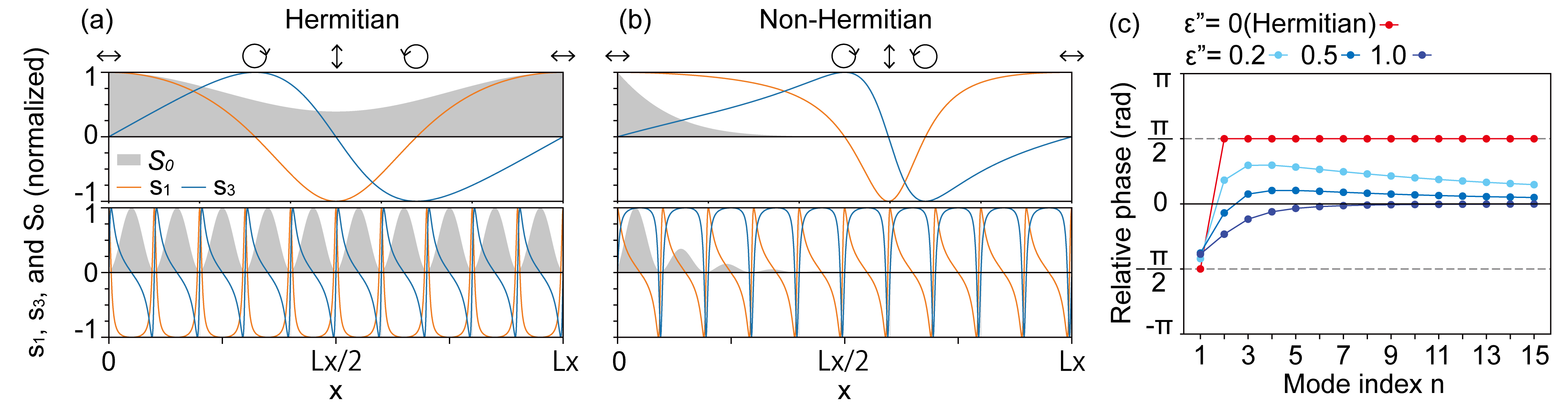}
\caption{Comparison of spatial distributions of Stokes parameters for (a) Hermitian and (b) non-Hermitian MT-symmetric cases. Top and bottom panels show the fundamental mode ($n=1$) and higher-order mode ($n=10$), respectively. Solid curves show the normalized Stokes parameters $s_1(x)$ (light blue) and $s_3(x)$ (orange), and the shaded gray regions show the intensity $S_0(x)$, normalized by its maximum value in each panel. The arrows above the $n=1$ panels indicate representative local polarization states of the in-plane electric field at selected positions along $x$. (c) Extracted phase difference $\Delta\phi=\phi_0-\phi_3$ between the oscillatory parts of $S_0$ and $S_3$, plotted as a function of the mode index $n$ for different values of $\varepsilon''$.}
\label{fig:Stokes_profiles}
\end{figure*}

The same field-shift picture also accounts for another feature seen in Fig.~\ref{Fig:PySz_compare}: the periodic sign reversal of $P_y$.
In the Hermitian limit, $E_x$ and $H_z$ are aligned in their transverse profiles and in phase in time, so $P_y$ remains unidirectional for all $x$. In the NHSE skin modes, however, only $E_x$ is shifted by the non-Hermitian anisotropy. This shift creates regions where the local contributions to $P_y$ acquire opposite signs, yielding periodic reversals of the axial energy flow within the skin-localized region.
By contrast, $P_x$ remains identically zero, because the non-Hermitian shift does not appear in $E_y$ and $H_z$, which therefore retain temporal quadrature, as seen from Eqs.~(\ref{eq:Ey_MT_final}) and~(\ref{eq:Hz_MT_final}).

\subsection{Normalized Stokes parameters and polarization state}
We next examine how the local polarization state is distributed relative to the field intensity.
In photonic-crystal waveguides and related nanophotonic platforms, where local circular polarization and polarization singularities play key roles in spin-dependent directional coupling and unidirectional excitation~\cite{Young15PRL,Sollner15NatNano,Dai25NatCom}, the spatial relation between the extrema of $S_3$ and the intensity maxima is of particular interest.
For this purpose, we characterize the in-plane polarization state of the TE skin modes using the Stokes parameters of the in-plane electric field. Since $E_x$ carries an overall factor of $i$ relative to $E_y$ [Eq.~\eqref{eq:Ex_MT_final}], $S_2$ is identically zero and the polarization state is fully captured by $(S_1,S_3)$.
We therefore use
\begin{align}
S_0 &= |E_x|^2+|E_y|^2,\\
S_1 &= |E_x|^2-|E_y|^2,\\
S_3 &= 2\,\Im\!\left(E_x^{*}E_y\right),
\end{align}
and introduce the normalized quantities
\begin{equation}
s_j(x)\equiv \frac{S_j(x)}{S_0(x)}.
\end{equation}

Figure~\ref{fig:Stokes_profiles} compares the transverse polarization textures of the Hermitian reference case and a representative non-Hermitian MT-symmetric case with $\varepsilon''=1$. The solid curves show the normalized Stokes parameters $s_1(x)$ and $s_3(x)$ for the fundamental mode ($n=1$) and for a higher-order mode ($n=10$), while the shaded gray region indicates the intensity profile $S_0(x)$, normalized by its maximum value in each panel.

In the Hermitian limit, where $\varepsilon''=0$, the TE mode forms an ordinary oscillatory interference field, so that the local Stokes vector traces a closed loop on the Poincar\'e sphere over one transverse period~\cite{Shevchenko19PRA,Peng17OE}. 
In typical Hermitian structured fields, the local circular polarization is proportional to the curl of the energy flow, whereas the regions of large spatial gradient do not generally coincide with the intensity maxima. Accordingly, regions where $|s_3|$ becomes large tend to appear where the intensity is relatively weak. This is also the case in Fig.~\ref{fig:Stokes_profiles}(a), where the high-intensity part of the field remains nearly linearly polarized.

By contrast, in the higher-order NHSE mode shown in Fig.~\ref{fig:Stokes_profiles}(b), the relation between $s_1$ and $s_3$ is nearly reversed from that in the Hermitian case, and the maxima of $S_0$ and $|s_3|$ occur at almost the same positions. 
This behavior can be understood in terms of the spatial shift of $E_x$, as shown in Fig.~\ref{Fig:EHfield}(c) and schematically illustrated in Fig.~\ref{fig:PyS3_profiles}(c). Because the non-Hermitian shift of $E_x$ increases its spatial overlap with $E_y$, stronger overlap enables the intensity maxima to occur in regions of large local circular polarization.

It should be noted, however, that the spatial alignment between the maxima of $S_0$ and the extrema of $S_3$ does not improve monotonically with increasing mode order. This behavior is clearly seen in the phase difference between the oscillatory parts of $S_0(x)$ and $S_3(x)$, shown in Fig.~\ref{fig:Stokes_profiles}(c) as a function of the mode index $n$.
The phase difference is evaluated from the oscillatory components extracted by fitting both $S_0(x)$ and $S_3(x)$ with a common exponentially localized oscillatory form. We define
\begin{equation}
\Delta\phi_{S_0}\equiv \phi_0-\phi_3,
\label{eq:relphase_S0}
\end{equation}
where $\phi_0$ and $\phi_3$ denote the phases of the fitted oscillatory components of $S_0$ and $S_3$, respectively. A smaller $|\Delta\phi_{S_0}|$ means a closer spatial alignment between the oscillatory maxima of $S_0$ and the oscillatory extrema of $S_3$.
The detailed fitting procedure is described in Appendix~\ref{app:fittfunc}.

In the Hermitian limit $\varepsilon''=0$ (red markers in Fig.~\ref{fig:Stokes_profiles}(c)), the relative phase is locked to $|\Delta\phi_{S_0}|\simeq \pi/2$, but its sign changes discontinuously when the dominant contribution to the oscillatory part of $S_0$ switches between the $E_x$- and $E_y$-derived terms. While the mode index at which this crossover occurs depends on the parameter set, in the present case it occurs between $n=1$ and $n=2$.
This behavior originates from the oscillatory interference structure in the Hermitian case. Since a linear combination of $\sin^2(\beta_n x)$ and $\cos^2(\beta_n x)$ with different coefficients can be written as
\begin{equation}
A\sin^2(\beta_n x)+B\cos^2(\beta_n x)
=\frac{A+B}{2}+\frac{B-A}{2}\cos(2\beta_n x),
\end{equation}
the sign of the oscillatory part is determined by $B-A$ and therefore changes discretely when the dominant contribution switches between the $E_x$- and $E_y$-derived terms. Accordingly, $\Delta\phi_{S_0}$ changes discontinuously in the Hermitian limit.

For finite $\varepsilon''$, the phase relation deviates from $|\Delta\phi_{S_0}|\simeq \pi/2$ because the non-Hermitian spatial shift of $E_x$ is superimposed on the oscillatory structure already present in the Hermitian limit. For smaller $\varepsilon''$, this shift remains moderate, so the discrete phase jump inherited from the Hermitian case still strongly affects the low-order modes. As a result, $\Delta\phi_{S_0}$ exhibits a broad maximum at relatively low mode orders before decreasing again as the mode index increases. For larger $\varepsilon''$, by contrast, the non-Hermitian shift becomes dominant and $\Delta\phi_{S_0}$ approaches zero in an almost monotonic manner.

These results suggest that the spatial alignment between intensity maxima and strong local circular polarization can be tuned by varying $\varepsilon''$ and the mode order. In particular, with an appropriate choice of $\varepsilon''$, such overlap can be achieved even in relatively low-order modes.
Non-Hermitian anisotropy shifts the oscillatory intensity maxima toward the $S_3$-dominated regions, suggesting a route to realizing waveguide modes in which strong field intensity and strong local circular polarization are simultaneously available at nearly the same positions.

\section{Loss-biased anisotropic media}
\begin{figure}[t]
\centering
\includegraphics[width=0.45\textwidth]{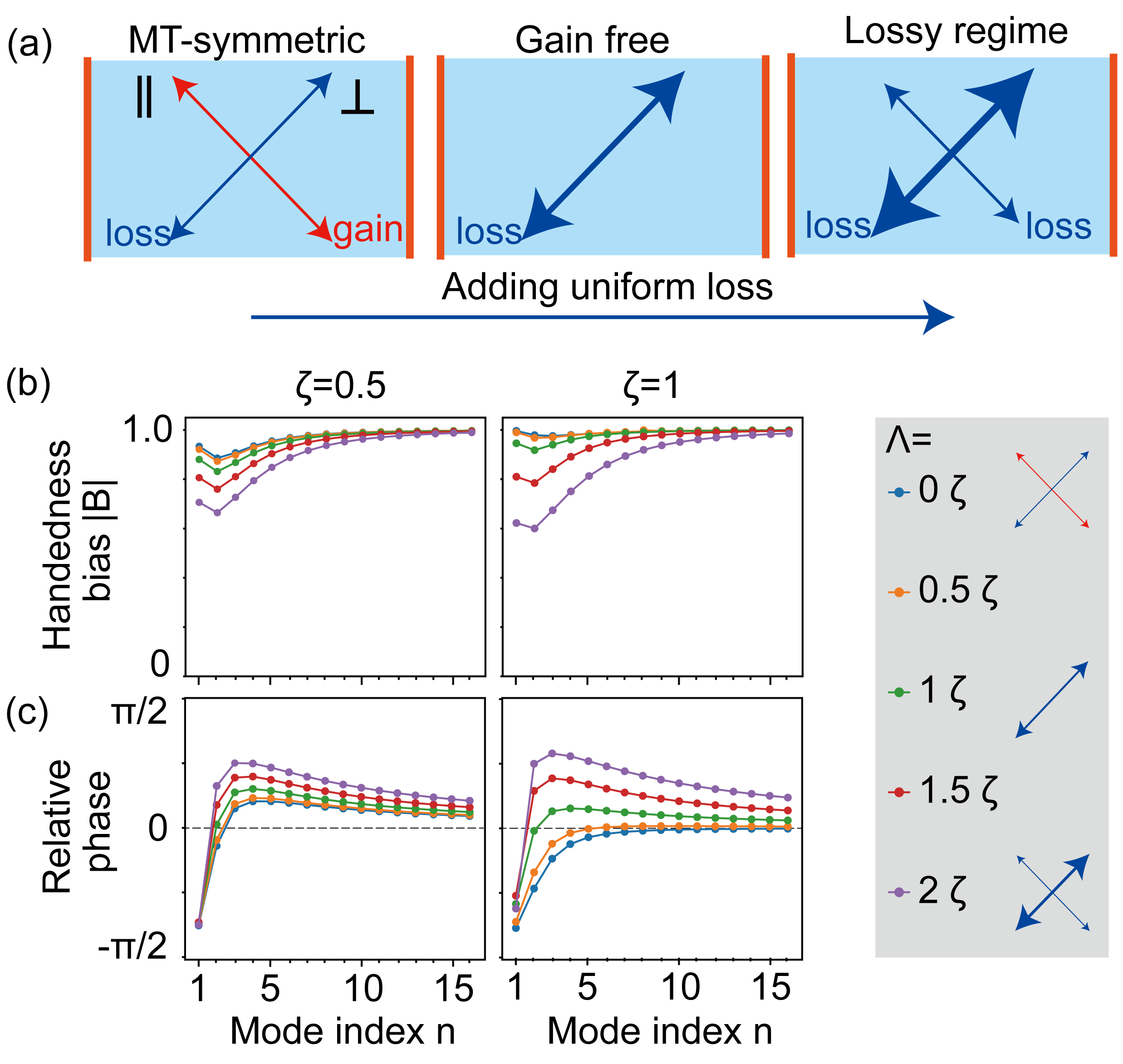}
\caption{(a) Schematic of the loss-biased medium. (b,c) Mode-index dependence of the (b) handedness-bias parameter $B$ and of the (c) extracted phase difference $\Delta\phi_{S_0}$. Here, $\zeta$ denotes the anisotropic gain/loss contrast between the principal axes of the underlying MT-symmetric medium, and $\Lambda$ is the added uniform loss offset.}
\label{fig:lossbiased}
\end{figure}

The analysis above focused on the MT-symmetric case, where compact closed-form expressions with purely real coefficient ratios are available. In practical photonic implementations, however, balanced gain and loss of the MT-symmetric form are often difficult to realize. Still, closely related skin modes can also emerge in loss-biased media, where both principal axes are lossy but with different loss rates~\cite{Takeda25PRA}.
Figure~\ref{fig:lossbiased}(a) schematically illustrates this configuration. Compared with the MT-symmetric case, the loss-biased medium can be viewed as a combination of a spatially uniform loss background and an anisotropic non-Hermitian contrast between the principal axes, so that the medium remains non-Hermitian and anisotropic without requiring optical gain.
We therefore examine whether the key polarization features identified in the MT-symmetric analysis remain in loss-biased anisotropic media. In particular, we examine whether the skin-localized modes still carry a finite $S_3$, and whether higher-order modes exhibit a noticeable handedness bias together with a tendency for the maxima of $|S_3|$ to coincide with the intensity maxima.

\subsection{Analytical calculation}
We model a loss-biased medium within the same rotated-anisotropy framework as Eq.~(\ref{eq:eps_rot45_general}).
The principal permittivities are taken as
\begin{equation}
\varepsilon_{\parallel}=a+i\gamma_{\parallel},\qquad
\varepsilon_{\perp}=a+i\gamma_{\perp},
\end{equation}
for which the in-plane permittivity tensor becomes
\begin{equation}
\varepsilon=
\begin{pmatrix}
a+i\bar\gamma & i\Delta\gamma\\
i\Delta\gamma & a+i\bar\gamma
\end{pmatrix},
\label{eq_eps_lossbiased}
\end{equation}
where
\begin{equation}
\bar\gamma=\frac{\gamma_{\parallel}+\gamma_{\perp}}{2},\qquad
\Delta\gamma=\frac{\gamma_{\parallel}-\gamma_{\perp}}{2}.
\end{equation}
Here, $\bar\gamma$ represents the spatially uniform loss background, while $\Delta\gamma$ gives the anisotropic non-Hermitian contrast that selects the skin-localization side. 

The general analytical solution in Eqs.~(\ref{eq:ori_Ey})--(\ref{eq:ori_Ex}) still applies to Eq.~(\ref{eq_eps_lossbiased}). 
In this case,
\begin{equation}
q=\frac{\varepsilon_{xy}+\varepsilon_{yx}}{2\varepsilon_{xx}}
=\frac{i\Delta\gamma}{a+i\bar\gamma}
\equiv q_r+i q_i,
\label{eq_q_lossbiased}
\end{equation}
so that the common factor becomes
\begin{equation}
e^{iqk_yx}=e^{-q_i k_y x}e^{i q_r k_y x}.
\end{equation}
The real part $q_r$ produces a phase slant along $x$ in the real-field distribution, whereas the imaginary part $q_i$ governs the skin-localized envelope.

To make the formal correspondence with the MT-symmetric case explicit, we define
\begin{equation}
\rho \equiv a+i\bar\gamma,\qquad
D \equiv \rho^2+(\Delta\gamma)^2,
\end{equation}
and rewrite the coefficients as
\begin{equation}
Z_x^\pm=\pm U,\qquad
Z_y^\pm=V\pm iW,
\label{eq:Z_lossbiased_compact}
\end{equation}
with
\begin{equation}
U = \frac{\beta \rho}{\omega\varepsilon_0 D},\qquad
V = \frac{k_y}{\omega\varepsilon_0 \rho},\qquad
W = \frac{\beta \Delta\gamma}{\omega\varepsilon_0 D}.
\end{equation}
Here $U$, $V$, and $W$ are generally complex, so Eq.~(\ref{eq:Z_lossbiased_compact}) should be understood as an algebraic reorganization rather than a decomposition into real and imaginary parts.

Using Eq.~(\ref{eq:Z_lossbiased_compact}), the field coefficients become
\begin{align}
B_C &= \frac{2i}{U},\qquad
B_S = 0,\nonumber\\
A_C &= \frac{2iV}{U}
= \frac{2ik_y D}{\beta \rho^2},\qquad
A_S = \frac{2iW}{U}
= \frac{2i\Delta\gamma}{\rho}.
\label{eq:AB_lossbiased}
\end{align}
Unlike in the MT-symmetric case, these coefficients are generally complex. As a result, the field components can no longer be reduced to the same purely real coefficient structure as in Eqs.~(\ref{eq:AB_MT})--(\ref{eq:Ex_MT_final}), and the strict quadrature relation between $E_x$ and $E_y$ does not generally hold. Nevertheless, the key structural features relevant to the present polarization texture remain. First, $B_S$ still vanishes identically, so $H_z$ retains a cosine-like profile along $x$. Second, $A_S$ is independent of the mode index and is determined only by the loss anisotropy through $\Delta\gamma/\rho$. Third, $A_C$ scales as $1/\beta\propto 1/n$, so the cosine component in $E_x$ becomes progressively weaker for higher-order modes. Thus, although the MT-symmetric relations are no longer exact, the same mode-dependent redistribution between the cosine- and sine-like contributions in $E_x$ survives in the loss-biased medium.

\subsection{FEM simulation}
We next examine these features by FEM eigenmode calculations while continuously varying the medium from the MT-symmetric gain-loss case to gain-free and purely lossy regimes. For this purpose, we add a spatially uniform loss offset $\Lambda$ to the MT-symmetric medium.
In the principal-axis basis, this is written as
\begin{equation}
\varepsilon_{\parallel}=(1+i\zeta)+i\Lambda,
\qquad
\varepsilon_{\perp}=(1-i\zeta)+i\Lambda,
\end{equation}
that is, $\gamma_{\parallel}=\zeta+\Lambda$ and $\gamma_{\perp}=-\zeta+\Lambda$. Thus, $\Lambda=0$ gives the MT-symmetric point, $\Lambda=\zeta$ the gain-free boundary, and $\Lambda>\zeta$ the purely lossy regime. In what follows, we examine $\zeta=0.5$ and $1.0$, and for each $\zeta$ sweep $\Lambda$ from $0$ to $2\zeta$ in steps of $0.5\zeta$.

First, we confirm whether the loss-biased medium still supports a finite $S_3$, and whether the mode-dependent handedness bias found in the MT-symmetric case remains. For this purpose, we quantify the balance between the positive and negative parts of the $S_3$ profile.
We integrate the positive part of $S_3(x)$ and the magnitude of its negative part over the full transverse range, and denote these by $P$ and $N$, respectively. Using these quantities, we define the bias parameter
\begin{equation}
B \equiv \frac{P-N}{P+N},
\label{eq:B_def}
\end{equation}
which satisfies $-1\le B\le 1$. A nonzero $|B|$ indicates an imbalance between the two handednesses of the in-plane circular polarization.

Figure~\ref{fig:lossbiased}(b) shows $|B|$ as a function of the mode index $n$. In all cases, $|B|$ remains large and generally approaches unity with increasing $n$, confirming that the handedness bias identified in the MT-symmetric analysis is retained also in loss-biased media.
The relatively large value already at the lowest mode reflects the fact that, in low-order skin modes, the negative part of $S_3$ appears only after substantial decay away from the localized edge, and therefore contributes little to the integral.

We next examine whether the tendency of the extrema of $S_3$ to approach the intensity maxima persists also in the loss-biased modes.
As in the MT-symmetric analysis above, we evaluate the phase difference $\Delta\phi_{S_0}$ between the oscillatory parts of $S_0$ and $S_3$ using the same fitting procedure as in Eq.~(\ref{eq:relphase_S0}). 
Figure~\ref{fig:lossbiased}(c) shows the extracted $\Delta\phi_{S_0}$ as a function of the mode index $n$.
The overall trend is qualitatively consistent with the MT-symmetric analysis, in the sense that stronger non-Hermitian anisotropy tends to reduce $|\Delta\phi_{S_0}|$ and hence to bring the extrema of $S_3$ closer to the intensity maxima.

Taken together, these results show that, although the detailed dependence on $\zeta$ and the uniform loss offset is non-universal, the key qualitative polarization features identified in the MT-symmetric analysis remain robust in loss-biased anisotropic media.

\section{Discussion and conclusions}
Using closed-form analytical solutions and FEM eigenmode calculations, we have shown that optical NHSE skin modes carry a finite transverse spin texture, represented by the in-plane circular-polarization texture of the electric field. We further found that this spin texture is qualitatively distinct from that of conventional lossless structured fields.
Analytical decomposition further shows that the gain/loss-induced exponential skin envelope does not generate an additional $S_3$ contribution, but acts as a common envelope modulating the oscillatory texture.
The familiar lossless relation between the energy-flow gradient and the local circular-polarization texture survives only for the oscillatory part, with an additional non-Hermitian offset. This offset accounts for the handedness bias in the full profile.
We also find that non-Hermiticity can bring the extrema of $S_3$ close to the intensity maxima. Furthermore, FEM calculations show that these characteristic polarization features remain qualitatively robust even in loss-biased anisotropic media beyond the ideal MT-symmetric limit.

While the present work has focused on the Stokes parameters and the energy flow as robust descriptors of local polarization, extending the present approach to a full spin-density description would provide a more complete picture of spin-dependent light--matter interactions in non-Hermitian anisotropic systems.

In addition, the same mechanism may be extended beyond the straight waveguide geometry to confined or curved configurations and to combined spin and orbital angular-momentum (SAM--OAM) physics.
In azimuthal NHSE structures~\cite{Takeda25PRA}, the transverse Stokes texture of a skin mode would map onto a closed trajectory along the edge, potentially enabling nontrivial spin textures together with orbital circulation.
SAM and OAM are widely used in nanophotonics for polarization-structured excitation, circularly polarized illumination, and optical vortex beams~\cite{Arikawa20SciAd,Tanaka23LPR,Ichiji24NanoP}. In this context, such extensions may provide additional routes for mode excitation, diagnostics, and spin-selective light--matter interactions in non-Hermitian photonic systems.

These results also suggest that polarization analysis can provide an additional handle for identifying and characterizing optical NHSE modes.
Conversely, the characteristic polarization textures of NHSE skin modes may be useful for designing excitation schemes in non-Hermitian photonic structures, particularly when local circular polarization and field intensity need to be controlled simultaneously.

More broadly, the present results show that lossy and non-Hermitian photonic media can host polarization textures that are not naturally obtained in conventional lossless interference structure or evanescent settings.
This expands the accessible design space of structured electromagnetic fields and provides a route toward more flexible control of local polarization and chirality in photonic systems with gain and loss.

\section*{Acknowledgments}
This work was supported by the JSPS KAKENHI (JP20H05641, JP24K01377, JP24H02232, JP24H00400, JP23KJ0355, JP25K17906).

\appendix
\renewcommand\thefigure{A\arabic{figure}} 
\setcounter{figure}{0} 

\section{Comparison with a conventional evanescent surface wave}
\label{app:spp_compare}
We briefly compare the NHSE skin mode with a conventional TM evanescent surface wave, such as a plane SPP, following the standard description of transverse spin in evanescent fields~\cite{Bliokh12PRA,Shi21PNAS,Ichiji23PRA}.
We consider a monochromatic TM evanescent wave propagating along $y$ and decaying along $x$,
\begin{equation}
\mathbf{E}=(E_x,E_y,0)e^{ik_y y-\kappa x},\quad
\mathbf{H}=(0,0,H_z)e^{ik_y y-\kappa x},
\end{equation}
where $k_y$ and $\kappa$ are real and positive.
For a lossless homogeneous medium, the transversality condition gives
\begin{equation}
E_x=i\frac{k_y}{\kappa}E_y .
\end{equation}

Thus, the normal and longitudinal electric-field components are in temporal quadrature.
At the same time, all field components share the same evanescent factor $e^{-\kappa x}$.
Their amplitudes are therefore maximal at the interface and decay monotonically away from it.
The third Stokes parameter in the $x$-$y$ plane then becomes
\begin{equation}
S_3=2\,\mathrm{Im}(E_x^*E_y)
=-2\frac{k_y}{\kappa}|E_y|^2
\propto e^{-2\kappa x},
\end{equation}
up to the sign determined by the propagation direction.

The time-averaged Poynting vector is
\begin{equation}
P_x=\frac{1}{2}\mathrm{Re}(E_yH_z^*)=0,\qquad
P_y=-\frac{1}{2}\mathrm{Re}(E_xH_z^*)\propto e^{-2\kappa x}.
\end{equation}
Therefore, in a conventional evanescent surface wave, the axial energy flow and the transverse circularity both follow the same evanescent decay away from the interface.

This behavior contrasts with the MT-symmetric NHSE skin mode discussed in the main text.
Although the NHSE skin mode also exhibits a surface-wave-like energy-flow structure with vanishing $P_x$ and boundary-localized $P_y$, its electric-field components are qualitatively different from those of a conventional surface wave.
The common exponential envelope acts as an amplitude mask, while the local polarization state is determined by the oscillatory factors: $E_y$ remains sine-like, $H_z$ remains cosine-like, and $E_x$ contains both cosine- and sine-like contributions.
Consequently, the transverse spin texture does not simply follow the monotonic evanescent envelope, but originates from the oscillatory component modified by the non-Hermitian shift of $E_x$.

\section{Comparison between $S_3$ and $\partial_xP_y$}
\label{app:dPy_compare}

\begin{figure}[t]
\centering
\includegraphics[width=0.48\textwidth]{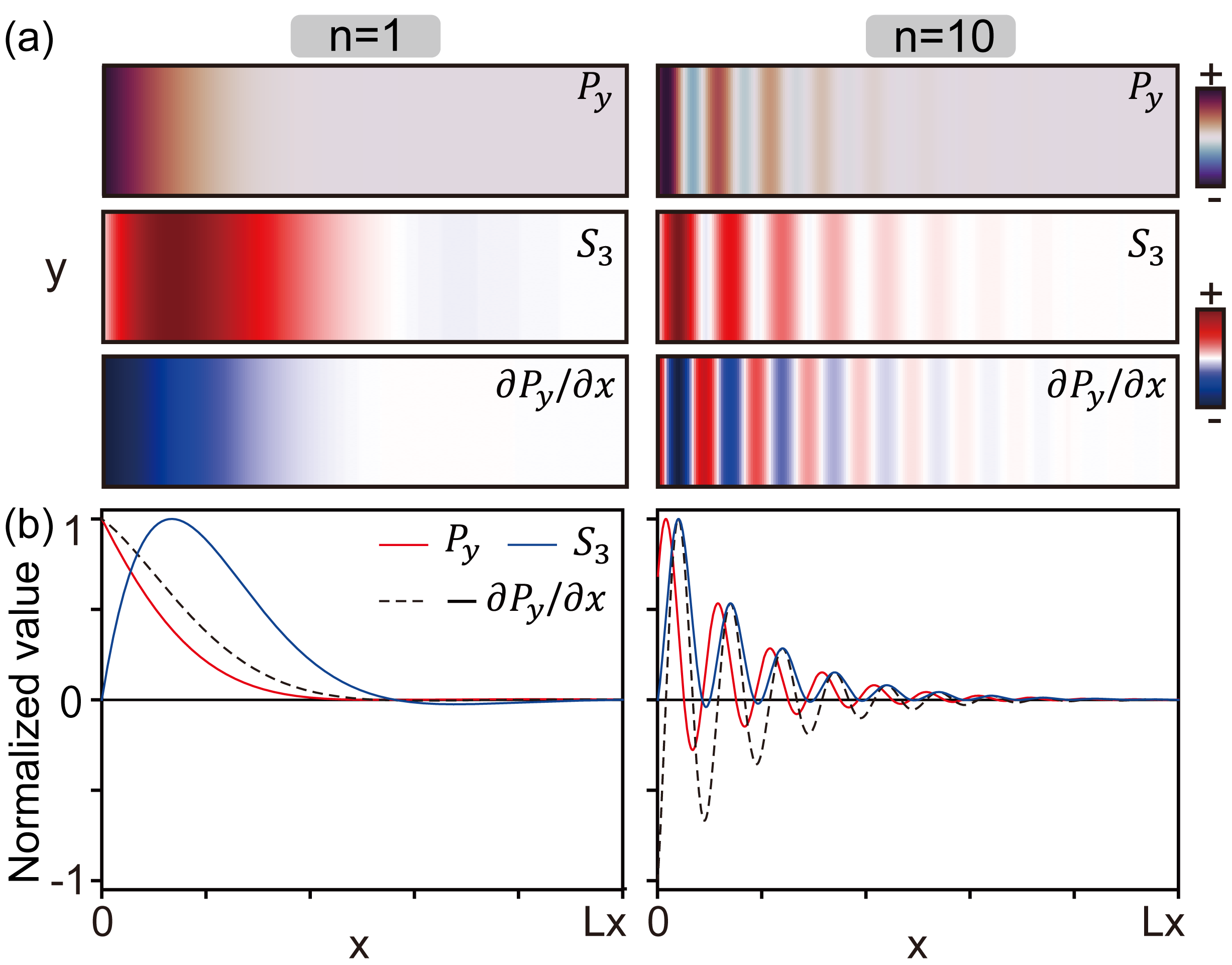}
\caption{Comparison between $S_3$ and the spatial derivative of the axial energy flow for MT-symmetric skin modes. (a) Two-dimensional maps of $P_y$, $S_3$, and $\partial_xP_y$ for $n=1$ and $n=10$. (b) Representative cross sections of $P_y$, $S_3$, and $-\partial_xP_y$, normalized by their maximum absolute values. The sign of the derivative is reversed only in panel (b) to facilitate comparison with the conventional lossless relation.}
\label{fig:dPy_compare}
\end{figure}

From Eq.~(\ref{eq:Py_MT}), the full derivative of the axial energy flow is
\begin{align}
\frac{\partial P_y}{\partial x}
=&\frac{e^{-2\alpha k_yx}}{4u(n)^2}
\Bigl[
-2\alpha k_y\{v(1+\cos2\beta x)+\alpha u(n)\sin2\beta x\} \nonumber\\
&\hspace{18mm}
-2\beta v\sin2\beta x
+2\beta\alpha u(n)\cos2\beta x
\Bigr].
\label{eq:dPy_full}
\end{align}

The first term in the square brackets originates from the derivative of the exponential skin envelope.
The remaining two terms correspond to the derivative of the oscillatory part of $P_y$; after multiplication by $-1/(2\beta)$ and addition of the offset $\alpha u(n)$, they reproduce the bracket of $S_3$ in Eq.~(\ref{eq:S3_MT}).

Figure~\ref{fig:dPy_compare}(a) compares two-dimensional distributions of $P_y(x)$, $S_3(x)$, and $\partial_xP_y(x)$ for representative MT-symmetric skin modes.
Panel (b) shows representative cross sections normalized by their maximum absolute values.
In panel (b), the derivative profile is plotted as $-\partial_xP_y$ so that its sign corresponds to the conventional lossless relation in the Hermitian limit.

For the fundamental mode, both $S_3$ and $-\partial_xP_y$ are localized near the boundary, but their profiles are already clearly different.
The derivative profile is largest at the boundary and decays almost monotonically, whereas $S_3$ vanishes at $x=0$ because $E_y=0$ and reaches its maximum at a finite distance from the boundary.
For the higher-order mode, the difference becomes more pronounced.
The full derivative $-\partial_xP_y$ exhibits a sign-changing oscillatory profile with a sizable negative part, reflecting the derivative of both the exponential envelope and the oscillatory modulation.
By contrast, $S_3$ shows a handedness-biased profile in which one sign is strongly suppressed.

These observations are consistent with the decomposition in Eqs.~(\ref{eq:dPy_full}) and~(\ref{eq:G_F_relation}): the observed $S_3$ profile is obtained by applying the differential relation to the oscillatory part, adding the non-Hermitian offset, and then multiplying by the common exponential envelope.
On this basis, we identify the envelope-derivative term in Eq.~(\ref{eq:dPy_full}) as an amplitude-gradient contribution to $\partial_xP_y$, rather than as a source term for $S_3$.

\section{Fitting function}
\label{app:fittfunc}
To quantify the relative positions of the oscillatory parts of $S_0$ and $S_3$, we extracted their phase difference from the analytic mode profiles at fixed $k_y$.
This procedure is nontrivial because both profiles consist of an exponentially localized envelope multiplied by a mode-dependent mixture of sine-like and cosine-like oscillations, together with a finite offset. As a result, the phase relation between $S_0$ and $S_3$ cannot be read off directly from the raw profiles in a unique manner, especially when the relative weights of the sine-like and cosine-like components vary with $\varepsilon''$ and with the mode order.

For each mode and each value of $\varepsilon''$, we therefore fitted the analytic profiles of $S_0(x)$ and $S_3(x)$ with a common exponentially localized oscillatory form,
\begin{equation}
\tilde S_j(x)\simeq e^{-\lambda x}\left[c_{j,0}+c_{j,c}\cos(k_x x)+c_{j,s}\sin(k_x x)\right],
\label{eq:fitfunction}
\end{equation}
where $j=0,3$, $\tilde S_0=S_0/\max_x S_0$, and $\tilde S_3=S_3/\max_x|S_3|$. Here, $\lambda$ represents the localization rate and $k_x$ is the effective transverse wavenumber of the oscillatory part. In the fitting procedure, $\lambda$ and $k_x$ were taken to be common to both $\tilde S_0$ and $\tilde S_3$, whereas the coefficients $c_{j,0}$, $c_{j,c}$, and $c_{j,s}$ were determined independently for each profile.

The oscillatory part of each fitted profile can then be written in the equivalent amplitude-phase form
\begin{equation}
c_{j,c}\cos(k_x x)+c_{j,s}\sin(k_x x)
=
A_j\cos\!\left(k_x x-\phi_j\right),
\end{equation}
with
\begin{equation}
A_j=\sqrt{c_{j,c}^2+c_{j,s}^2},
\qquad
\phi_j=\arg(c_{j,c}+i c_{j,s}).
\end{equation}
Using these fitted phases, we defined the relative phase difference between the oscillatory parts of $S_0$ and $S_3$ as
\begin{equation}
\Delta\phi=\phi_0-\phi_3.
\end{equation}

This definition isolates the phase relation of the oscillatory components from the overall exponential localization and from the offset term. It therefore provides a convenient measure of how the circular-polarization extrema are positioned relative to the oscillatory part of the intensity profile. The values plotted in Fig.~\ref{fig:Stokes_profiles}(c) were obtained in this manner.

\bibliography{sorsamp}

\end{document}